\documentclass[11pt]{article}
\usepackage{amssymb,amsmath,graphicx,url,amsmath,hyperref}
\usepackage[top=2.5 cm,right=3cm,bottom=2cm,left=3cm]{geometry} 
\graphicspath{{figures/}}
\def\isdef{\mbox {$\ \stackrel{\rm def}{=} \ $}}
\newtheorem{precor}{{\bf Corollary}}

\newtheorem{precon}{{\bf Conjecture}}

\newtheorem{predefin}{{\bf Definition}}

\newtheorem{preexm}{{\bf Example}}

\newtheorem{preappl}{{\bf Application}}

\newtheorem{prelem}{{\bf Lemma}}

\newtheorem{preproof}{{\bf Proof.\ }}

\newtheorem{presproof}{{\bf Sketch of Proof.\ }}

\newtheorem{prethm}{{\bf Theorem}}

\newtheorem{prealphthm}{{\bf Theorem}}

\newenvironment{alphthm}{\begin{prealphthm}{\hspace{-0.5 em}{\bf.\ }}}{\end{prealphthm}}
\newtheorem{prepro}{{\bf Proposition}}

\newtheorem{preprb}{{\bf Problem}}

\def\isdef{\mbox {$\ \stackrel{\rm def}{=} \ $}}

\numberwithin{equation}{subsection}
\def\conct[#1,#2]{\mbox {${#1} \leftrightarrow {#2}$}}
\def\dconct[#1,#2]{\mbox {${#1} \rightarrow {#2}$}}
\def\deg[#1,#2]{\mbox {$d_{_{#1}}(#2)$}}
\def\mindeg[#1]{\mbox {$\delta_{_{#1}}$}}
\def\maxdeg[#1]{\mbox {$\Delta_{_{#1}}$}}
\def\outdeg[#1,#2]{\mbox {$d_{_{#1}}^{^+}(#2)$}}
\def\minoutdeg[#1]{\mbox {$\delta_{_{#1}}^{^+}$}}
\def\maxoutdeg[#1]{\mbox {$\Delta_{_{#1}}^{^+}$}}
\def\indeg[#1,#2]{\mbox {$d_{_{#1}}^{^-}(#2)$}}
\def\minindeg[#1]{\mbox {$\delta_{_{#1}}^{^-}$}}
\def\maxindeg[#1]{\mbox {$\Delta_{_{#1}}^{^-}$}}
\def\isdef{\mbox {$\ \stackrel{\rm def}{=} \ $}}
\def\dre[#1,#2,#3]{\mbox {${\cal E}_{_{#3}}(#1,#2)$}}
\def\pdre[#1,#2,#3]{\mbox {${\cal P}_{_{#3}}(#1,#2)$}}
\def\var[#1,#2]{\mbox {${\rm Var}_{_{#1}}(#2)$}}
\def\ls[#1]{\mbox {$\xi^{^{#1}}$}}
\def\hom[#1,#2]{\mbox {${\rm Hom}({#1},{#2})$}}
\def\onvhom[#1,#2]{\mbox {${\rm Hom^{v}}(#1,#2)$}}
\def\onehom[#1,#2]{\mbox {${\rm Hom^{e}}(#1,#2)$}}
\def\core[#1]{\mbox {$#1^{^{\bullet}}$}}
\def\cay[#1,#2]{\mbox {${\rm Cay}({#1},{#2})$}}
\def\cays[#1,#2]{\mbox {${\rm Cay_{s}}({#1},{#2})$}}
\def\dirc[#1]{\mbox {$\stackrel{\rightarrow}{C}_{_{#1}}$}}
\def\cycl[#1]{\mbox {${\bf Z}_{_{#1}}$}}

\def\sdg[#1]{\mbox {$\stackrel{\leftrightarrow}{#1}$}}

\newcommand{\Integer}{\mathbb{N}}

\newcommand{\Real}{\mathbb{R}}

\newcommand{\Field}{\mathbb{F}}

\newcommand{\calM}{{\cal M}}
 
\newcommand{\outputs}{{\rightarrow}}

\newcommand{\cpsc}{ PLCIE }

\newcommand{\state}{\mathbf{s}}

\newcommand{\keystream}{z}
\newcommand{\plain}{p}
\newcommand{\cipher}{c}
\newcommand{\keystreamvec}{\mathbf{z}}
\newcommand{\plainvec}{\mathbf{p}}
\newcommand{\ciphervec}{\mathbf{c}}
\newcommand{\errorve}{\mathbf{e}}
\newcommand{\mem}{\tilde{c}}
\newcommand{\memvec}{\mathbf{\tilde{c}}}
\newcommand{\key}{k}
\newcommand{\len}{\ell}

\newcommand{\Wmat}{\mathbf{W}}

\newcommand{\Bmat}{\mathbf{B}}
\newcommand{\Fmat}{\mathbf{F}}
\newcommand{\Dmat}{\mathbf{D}}
\newcommand{\Emat}{\mathbf{E}}

\newcommand{\Amat}{\mathbf{A}}
\newcommand{\I}{\mathbf{I}}

\newcommand{\memM}{\mathbf{M}}
\newcommand{\BigPi}{\wp}
\newcommand{\keyspace}{{\cal K}}
\newcommand{\nextstatefunc}{\varphi}
\newcommand{\keystreamgen}{\gamma}

\newcommand{\encryptionfun}{\varepsilon}
\newcommand{\decryptionfun}{\delta}
\newcommand{\kernel}{\mathrm{Kernel}}

\newcommand{\transmitter}{\mathrm{Enc}}
\newcommand{\receptor}{\mathrm{Dec}}

\begin{document}
\begin{center}

{\Large \bf  A Self-synchronized Image Encryption Scheme}\\
\vspace*{0.5cm}
{\bf Amir Daneshgar
\footnote{Correspondence should be addressed to {\tt daneshgar@sharif.ir}.
}
 and Behrooz Khadem}\\
{\it Sharif University of Technology - Department of Mathematical Sciences} \\
{\it P.O. Box {\rm 11155--9415}, Tehran, Iran.}\\
{\tt  daneshgar@sharif.ir }\\ 
{\it Kharazmi University - Faculty of Mathematics and Computer Science} \\
{\it P.O. Box {\rm 15719-14911}, Tehran, Iran.}\\
{\tt std$_{-}$khadem@khu.ac.ir} \\
\end{center}
\begin{abstract}
\noindent
In this paper, a word based chaotic image encryption scheme for gray images is proposed, that can be used in both 
synchronous and self-synchronous modes.
The encryption scheme operates in a finite field where we have also analyzed its performance according to numerical precision
used in implementation. We show that the scheme not only passes a variety of security tests, but also it is 
verified that the proposed scheme operates faster than other existing schemes of the same type even 
when using lightweight  short key sizes. \\

$\textbf{\textsf{KeyWords: chaos, image encryption, self-synchronization.}}$
\end{abstract}
\section{Introduction}

Several image encryption schemes have been proposed in the literature based on different approaches for design or implementation,
while chaos-based encryption schemes have the advantage of presenting a good combination of speed and security.

It seems that Fridrich \cite{Fr97} is among the first contributors who has proposed an image encryption scheme  based on chaotic maps, where in \cite{Fr97} certain invertible chaotic 2D maps on a torus or on a square have been used to create new symmetric block encryption schemes. Many other chaotic image encryption schemes have been proposed ever since with different properties and 
motivations for application (e.g. see \cite{Fra14,GHR12,KT07,MGR11,SK10} and references therein).  

Strictly speaking, one may consider the following challenges when one is trying to design an image encryption scheme
(see \cite{MD11,SK10,SZJ12} and references therein):

\begin{itemize}
\item{The scheme must have a relatively high speed of performance since images usually consist of large blocks of data. }
\item{Since the information content of an image is contained in high frequencies the scheme must possess a high mixing performance.}
\item{According to typical applications, the scheme must be relatively lightweight and should be able to operate with relatively small keys with acceptable security guaranties.}
\item{The scheme must guaranty secure, reliable and fast rates of data transfer. }
\end{itemize}

Considering above facts, chaos-based stream ciphers may seem to be a solution while,

\begin{itemize}

\item{Although, concentrating on chaotic word-based designs operating in a finite field may seem to be a solution for fast and reliable encryption, one must note that discretizing chaotic maps usually deteriorate their chaotic properties that may lead to weak security conditions.}
\item{Data transfer reliability can be achieved using self-synchronization, however, security guaranty is much harder in presence of self-synchronization for the feedback structure.}
\end{itemize}
It seems that one of the main problems with chaotic encryption schemes introduced so far is the direct application of the chaotic
sequence which is far from being pseudorandom when it is digitized, which will definitely lead to security weaknesses when the scheme is not design properly (e.g. see \cite{LL09,TXM08}).
Therefore, to solve the above mentioned and seemingly contradicting challenges, we introduce an image encryption scheme in which we
have used a chaotic string indirectly to generate a pseudorandom permutation whose pseudorandomness is guaranteed by the results of \cite{AFR11}. On the other hand to compensate the weakness of discrete permutations in uniformly encrypting the high frequency 
image data (mainly based on correlation along edges) we use a linear feedback to achieve the acceptable uniformization. 
In other words we,
\begin{itemize}
\item{Use pseudorandom permutations generated by chaotic maps.}
\item{Use word-based chaos to guaranty fast encryption.}
\item{Compensate discretization phenomenon using a fast linear feedback.}
\item{Make sure that the scheme can perform in both synchronous and self-synchronous modes by setting parameters, to be able 
to be used in different channel conditions in a reliable way.}
\item{Make sure that the scheme has a fast receiver as an unknown input observer of the transmitter.}
\end{itemize}

\cpsc \footnote{PseudoLinear Chaotic Image Encryption (also see \cite{DM?} for a switching version and its properties).} is an extension of {\rm PLC} scheme introduced in \cite{KDM13} tuned to be used for image encryption. PLCIE is a word-based chaotic encryption scheme having $ \ell$-word state vectors that can be controlled by users, giving sufficient flexibility for multi-level security.
\cpsc consists of a Initializing phase, internal state update, memory update, encryption and decryption that will be described in detail in Section~\ref{plc scheme}.
 In Section~\ref{analysis}, we apply various tests to verify the performance and the security of the proposed scheme. 

\section{Description of \cpsc} \label{plc scheme}

A digital image usually can be interpreted as a function
  $ z=f(x,y) $ 
of physical horizontal
 $  x$ 
and vertical 
 $  y$
 coordinates, that determine illumination or grayscale value of the picture element (or the pixel) at location
 $ (x,y) $.
A pixel is the smallest addressable element in a display device. The level of illumination at each pixel has a value between 0 and 255. Thus, in a digital image, the grayscale of each pixel is presented by one byte and the whole image is presented by a large matrix of bytes. 
The histogram of a digital image 
is a discrete function 
$ h (r_{_{k}}) = n_{_{k}} $, where  
$ r_{_{k}} $
 is the 
 $k$-th 
gray level and 
$ n_{_{k}}$
 is the number of pixels of the image with gray level
 $ r_{_{k}} $.

Let $q$ be a prime power, $\Field_{_{q}}$ be the finite field on $q$ elements\footnote{The cryptosystem can be defined on any finite field, however,  in our real life applications with a lightweight setup  we set $\Field_{_{q}}=GF(16)$ or $\Field_{_{q}}=GF(17)$.}
and  $ f :\Real \outputs \Real$
 be a chaotic map (e.g. as in \cite{De03}). Consider a family of maps as
 $\pi: \keyspace \times \Field_{_{q}} \outputs \Field_{_{q}} $
  such that for any 
$\key \in \keyspace$
 the map 
$ \pi(\key,.) $
 is a discrete chaotic permutation on
  $ \Field_{_{q}} $ as a
  discrete approximation of 
 $ f $ 
(e.g. as defined in \cite{AKS07}). The two-variable map
 $ \pi $
 gets a value
 $\key \in \keyspace$
 as well as a field element
 $a\in \Field_{_{q}} $, and returns
 $\pi_{_{\key}}(a) \isdef \pi(\key,a)$. 
In \ref{init phase} we will describe how one may compute this family of chaotic permutations.

For all   
$ t\geq1 $, consider 
$ \plain_{_{t}}, \cipher_{_{t}}, \keystream_{_{t}} \in \Field_{_{q}} $,  and
let $ \left< \plain_{_{t}}\right>$, $ \left< \cipher_{_{t}}\right> $, $\left< \keystream_{_{t}}\right> $
 be the plain, the cipher, and the keystream sequences in time, respectively.
Let $ \len $ be an integer. Also, define column vectors  $ \plainvec_{_{t}},\ciphervec_{_{t}},\keystreamvec_{_{t}} $
in $\Field_{_{q}}^{^\len}$ as
\begin{center}
$\plainvec_{_{t}}\ \isdef [\plain_{_{t}}^{^{(1)}},0, \cdots,0]^{^{T}}  , \ \  \ciphervec_{_{t}}\ \isdef [\cipher_{_{t}}^{^{(1)}},\cipher_{_{t}}^{^{(2)}}, \cdots, \cipher_{_{t}}^{^{(\ell)}}]^{^{T}},
\ \ \keystreamvec_{_{t}} \isdef [\keystream_{_{t}}^{^{(1)}},\keystream_{_{t}}^{^{(2)}},\cdots,\keystream_{_{t}}^{^{(\ell)}}]^{^{T}} . $
\end{center}
The internal state
 $\state_{_{t}} \in \Field_{_{q}}^{^{\len}}$ 
and internal memory
  $\memvec_{_{t}} \in \Field_{_{q}}^{^{\len}}$  
are also defined as column vectors   

\begin{center}
$\state_{_{t}} \isdef [s_{_{t}}^{^{(1)}},s_{_{t}}^{^{(2)}}, \cdots, s_{_{t}}^{^{(\len)}}]^{^{T}} , \quad  \memvec_{_{t}} \isdef [\mem_{_{t}}^{^{(\len)}},\mem_{_{t}}^{^{(\len -1)}}, \cdots, \mem_{_{t}}^{^{(1)}}]^{^{T}} . $
\end{center}

 Define the map
 $ \BigPi_{_{\key}}:\Field^{^{\len}}_{_{q}} \outputs \Field^{^{\len}}_{_{q}}$
 as follows

\begin{center}
$\BigPi_{_{\key}}([a_{_{1}},a_{_{2}}, \cdots, a_{_{\len}}]^{^{T}})\ \isdef [\pi_{_{\key}}(a_{_{1}}),\pi_{_{\key}}(a_{_{2}}), \cdots, \pi_{_{\key}}(a_{_{\len}})]^{^{T}}. $
\end{center}
\cpsc scheme uses a set of functions introduced in Table \ref{map-table} in which $ \calM $ stands for the set of all
  $ \len \times \len $ matrices on $ \Field_{_{q}}$. 
Also \cpsc has a initializing phase along with two other main phases called the 
 kernel computation phase, and the encryption/decryption phase that will be described in what follows.

\begin{table}[!t]
\renewcommand{\arraystretch}{1.3}
\caption{Functions used in PLCIE}
\centering
\label{map-table}
\begin{tabular}{|c|c|}
\hline Title& Form \\
\hline 
\hline State update& $ \nextstatefunc_{_{\key}}: (\Field^{^{\len}}_{_{q}})^{^{2}} \times \calM^{^{4}} \outputs \Field^{^{\len}}_{_{q}}$ \\
\hline Keystream generator& $ \keystreamgen_{_{\key}}: (\Field^{^{\len}}_{_{q}})^{^{2}}\times \calM ^{^{2}} \outputs \Field^{^{\len}}_{_{q}}$ \\
\hline Encryption& $\encryptionfun_{_{\key}}: (\Field^{^{\len}}_{_{q}})^{^{2}}\times \calM \outputs \Field^{^{\len}}_{_{q}}$ \\ 
\hline  Decryption& $\decryptionfun_{_{\key}}: (\Field^{^{\len}}_{_{q}})^{^{2}}\times \calM \outputs \Field^{^{\len}}_{_{q}}$ \\
\hline  Memory update& $\mu: (\Field^{^{\len}}_{_{q}})^{^{2}}\outputs \Field^{^{\len}}_{_{q}}$ \\
\hline
\end{tabular}
\end{table}

\subsection{The initializing phase}\label{init phase}
In this phase, a chaotic sequence is produced, that gives rise to the pseudorandom permutation $\pi_{_{\key}}$. 
Also, the initial value vector $IV$
is set according to a uniform distribution. The secret key is a binary string consisting of 
\begin{itemize}
\item{encoding of the system precision $ prec $ (one bit indicating $16$ or $32$ bits representation of numbers).}
\item{encodings of the initial values of the chaotic map  ($ r_{_{0}},l_{_{0}}$), chosen uniformly at random, where
$ r_{_{0}} \in_{_{R}} (0,1) $ (presented in $prec$ bits floating point format) and $ l_{_{0}} \in_{_{R}} \{1,2,\cdots, 2^{^{prec}}-1\} $ (presented in $ prec $ bits integer format)}.
\item{encodings of a number $ a \in_{_{R}} \Field_{_{q}} $  and 
$(i_{_{1}}, j_{_{1}}, e_{_{i_{_{1}}j_{_{1}}}}), \cdots, (i_{_{n}}, j_{_{n}}, e_{_{i_{_{n}}j_{_{n}}}})$,
in which $n < \frac{l^2}{2}$,
indicating an encoding of the matrix $\Emat$ (to be used later) such that the entries not mentioned in the coding is set to the default value $a$. }

\end{itemize}
Let $\iota$ be a constant integer of order $O(\len)$ (e.g. for $\len=8$ this parameter can be chosen as $\iota=32$).
Then,  $ IV $  is a $2 \ell $  word vector  which is used to preset $ \state_{_{-\iota}} $  and $ \memvec_{_{-\iota}} $.
Note that here one may use a random string of length $\iota$ as a prefix of plaintext for whitening.
 
For the chaotic map we have chosen  a particular version of the R\'{e}nyi map \cite{AFR11} with parameter $ \beta=3 $ which is defined as follows,

\begin{equation}\label{psi}
\psi(x)=3 x - \lfloor 3 x \rfloor  \ \ \ ,x \in (0,1).
\end{equation}

In \cite{AFR11} the discrete version of this map (called $ \psi_{_{d}}(x) $) is defined as,
\begin{equation}\label{psi-d}
g(x)=\dfrac{\lfloor 2^{^{prec}}x\rfloor}{2^{^{prec}}}   \ \ , \ \
\psi_{_{d}}(x)=3 g(x) - \lfloor 3 g(x) \rfloor  \ \ \ ,x \in (0,1).
\end{equation}

  Also it is shown in the same reference that the map has a positive Lyapunov exponent and any of its' successive iterations has acceptable statistical properties. However, we will see that this map by itself is not good enough to be solely used in an image encryption scheme as a pseudorandom source for permutations (see Section~\ref{init phase}).
 
After producing the iterated sequence
    $ \{\psi_{_{d}}^{^{t}}(r_{_{0}})\}_{_{t = 1}}^{^{l_{_{0}}+q}} $, 
and eliminating the first 
 $ l_{_{0}} $
  transient elements, the sequence is used to generate a pseudorandom  permutation.
\begin{figure}[h]
\centerline{\includegraphics[width=5cm]{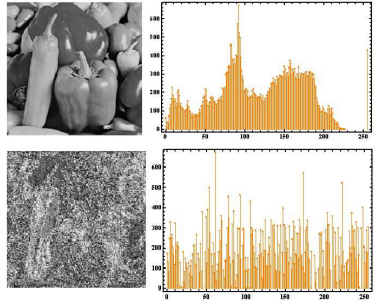}}
\caption{\label{fig:Only Chaos}   \ Peper original  versus  value-permuted  image.}
\end{figure}

The initializing algorithm gets the secret key  $ \key $,
 initial value $ IV $,  sequence\footnote{Here we may assume that the vector $\textbf{v}_{_{t}}$ does not have repeated entries 
 by chaotic properties of the Renyi map. Clearly since the probability of having a bad vector with two identical entries is negligible one may eventually find a good vector by repeating the process. }
  $\textbf{v}_{_{t}}= \{\psi_{_{d}}^{^{t}}(r_{_{0}})\}_{_{t =l_{_{0}}+1}}^{^{l_{_{0}}+q}} $
and  its sorted sequence $ \textbf{x}_{_{i}}= \{\psi_{_{d}}^{^{i}}(r_{_{0}})\}_{_{i=1}}^{^{q}} $,  and returns
 a key permutation $ \pi_{_{\key}} $,  an initial state vector $ \state_{_{-\iota}}$ and an initial memory vector
 $ \memvec_{_{-\iota}}$.
 To generate the chaotic permutation $ \pi_{_{\key}}$, 
we follow \cite{ASK05} and define,
\begin{equation}\label{Pi_k}
\pi_{_{k}}(i)=\ \left
\{
\begin{array}
{lll}
j&\ \,\ \ \psi_{_{d}}(x_{_{i}})=x_{_{j}} \\
j'&\ \,\ \ (v_{_{q-1}}=x_{_{i}}  \ and \  v_{_{0}} =x_{_{j'}}).
\end{array}
\right.
\end{equation}
Unfortunately, this pseudorandom permutation is not by itself sufficient for image encryption mainly because of correlations existing in an image, most of which concentrated in high frequency components (e.g. see Figure~\ref{fig:Only Chaos}).

\subsection{The kernel computation phase}  \label{ker phase}
This phase contains state update algorithm and key stream generator as shown in Equations  (\ref{kernel-yek2-1}) , (\ref{kernel-yek3-1}) and (\ref{kernel-yek4-1}).
In the next paragraph, synchronous and self-synchronous modes are explained. As one may note, both modes have similar chaotic maps, internal states and permutation-substitution components, but the self-synchronous mode has an internal memory and a memory update function in order to synchronize both transmitter and receiver simultaneously.
The  memory update function $ \mu $ is defined as follows,
$$\mu(\memvec_{_{t}},\ciphervec_{_{t-1}}) \isdef \memM \memvec_{_{t}}+\ciphervec_{_{t-1}}$$
in which
 \begin{center} \label{mem dfn}
$\memvec_{_{t}} = [\mem_{_{t}}^{^{(\len)}},\mem_{_{t}}^{^{(\len-1)}}, \cdots, \mem_{_{t}}^{^{(2)}}, \mem_{_{t}}^{^{(1)}}]^{^{T}} $, \ \ \
$ \mem_{_{t}}^{^{(i)}}=\cipher_{_{t-\len+i-1}} $, \ \ \

\ \ \

$\mathbf{M} \isdef \left [\begin{array} {llllll}
0&0&0&\cdots&0&0\\0&0&0&\cdots&1&0\\ \vdots&\vdots&\vdots&&\vdots&\vdots\\0&0&1&\cdots&0&0\\0&1&0&\cdots&0&0\\\end{array} \right]. $ 
\end{center}
The state update algorithm gets a current state
 $ \state_{_{t}} $
 and returns the next state $ \state_{_{t+1}}.$
 Clearly, as in  (\ref{kernel-yek2-1}), the state update algorithm consists of a linear part and a chaotic permutation. The linear part not only connects a relation between the internal state and previous cipher symbols, but also increases shuffling property which results in an almost perfect  uniformly distributed output. 
  The keystream generator algorithm gets the current state
  $ \state_{_{t}}  $
 and returns the keystream
 $ \keystreamvec_{_{t}}. $
 
 Let $ \Amat, \Bmat, \Dmat, \Emat, \Fmat, \Wmat \in \calM$
  while
 $ \Fmat $
 is invertible. 
  The kernel of  \cpsc  in the synchronous mode, $ \kernel_{_{s}}, $ is defined as follows.
\begin{equation}\label{kernel-yek2-1}
\kernel_{_{s}}:\left\lbrace
\begin{array}{llllll}
initial &:\pi_{_{\key}},\Amat,\Bmat,\Dmat,\\
\state_{_{t+1}}&=\Dmat \state_{_{t}}+\Amat\BigPi_{_{\key}}(\state_{_{t}})\\
\keystreamvec_{_{t}}&=\Bmat\BigPi_{_{\key}}(\state_{_{t}})\\
\end{array}\right. 
\end{equation}
To  control the error diffusion rate, correct state recovery and maintain stability, one may improve the above kernel to work in 
a self-synchronized mode as follows,
\begin{equation}\label{kernel-yek3-1}
E\kernel_{_{ss}}:\left\lbrace
\begin{array}{llllll}
initial &:\pi_{_{\key}},\Amat,\Bmat,\Dmat,\Emat, \Wmat \\
\state_{_{t+1}}&=\Wmat\memvec_{_{t}}+\Dmat \state_{_{t}}+\Amat\BigPi_{_{\key}}(\state_{_{t}})+\Emat\BigPi_{_{\key}}(\plain_{_{t}})\\
\memvec_{_{t+1}}&=\memM \memvec_{_{t}}+\ciphervec_{_{t-1}}  \\
\keystreamvec_{_{t}}&=\Wmat\memvec_{_{t}}+\Bmat\BigPi_{_{\key}}(\state_{_{t}}),\\
\end{array}\right. 
\end{equation}
Note that one get the synchronous mode when $\Wmat=\Emat=\bf{0}$. 
It is proved in \cite{KDM13} (see Theorem~\ref{thmsynch} below) that if
\begin{itemize}
\item[1-]
$\Amat=\Emat\Fmat^{-1} \Bmat $,
\item[2-]
there exits  $n_{_{0}}\in \Integer$ 
such that
$ \Dmat^{^{n_{_{0}}}}=\mathbf{\bf{0}}$
\end{itemize} 
 then $E\kernel_{_{ss}}$ has an (unknown input) observer defined as 
\begin{equation}\label{kernel-yek4-1}
D\kernel_{_{ss}} :
\left\lbrace 
\begin{array}{llllll}
initial &:\pi_{_{\key}},\left\langle  \ciphervec_{_{t}} \right\rangle ,\Amat,\Bmat,\Dmat,\Emat,\Fmat,\Wmat \\
\widehat{\state}_{_{t+1}}&=\Wmat\memvec_{_{t}}+\Dmat \widehat{\state}_{_{t}}+\Amat\BigPi_{_{\key}}(\widehat{\state}_{_{t}})+\Emat\Fmat^{^{-1}}(\ciphervec_{_{t}}-\widehat{\keystreamvec}_{_{t}})\\
\memvec_{_{t+1}}&=\memM \memvec_{_{t}}+\ciphervec_{_{t-1}}\\
\widehat{\keystreamvec}_{_{t}}&= \Wmat\memvec_{_{t}}+\Bmat\BigPi_{_{\key}}(\widehat{\state}_{_{t}}).\\
\end{array}
\right. 
\end{equation}
\subsection{The encryption/decryption phase}  \label{enc dec phase}
Based on the kernel equations we have the following procedures for encryption and decryption in general,
\begin{equation}\label{yek3-1}
\transmitter: \left\lbrace
\begin{array}{lll}
input:\left< \plain_{_{t}} \right>\\
E\kernel\\
output:\ciphervec_{_{t}}=\keystreamvec_{_{t}}+\Fmat\BigPi_{_{\key}}(\plainvec_{_{t}}).\\
\end{array}\right.
\end{equation}
\begin{equation}\label{yek4-2}
\receptor:\left\lbrace
\begin{array}{llll}
input:\left< \cipher_{_{t}} \right>\\
D\kernel \\
output:\widehat{\plainvec}_{_{t}}=\BigPi_{_{\key}}^{^{-1}}(\Fmat^{^{-1}}(\ciphervec_{_{t}}-\widehat{\keystreamvec}_{_{t}})).\\
\end{array}\right.
\end{equation}
Note that in this setting the matrix $\Emat$ is secret and is included in the key. The vector $IV$ is chosen at random and is sent along with the ciphertext to make sure that a trivial CPA attack is not applicable. 
Moreover, if one is not working in a lightweight setting then one may also encode the matrix $\Bmat$ in the key and make it secret to enhance security conditions of the scheme.

Based on the following theorem \cite{KDM13}, by making $\Amat$ secret as a function of the key and choosing $\Dmat$ properly, 
one may prove that a receiver as an UIO exists.
We recall the result along with a sketch of proof for the scheme as follows.
\begin{alphthm}\label{thmsynch} 
In {\rm \cpsc}, if 
\begin{itemize}
\item[{\rm a)}]
$\bf{A}=\bf{E}\bf{F}^{^{-1}}\bf{B}$,
\item[{\rm b)}]
The matrix $\Dmat$ is nilpotent i.e. there exists an integer $ n_{_{0}} \in \Integer $ such that $ \Dmat^{^{n_{_{0}}}}=\mathbf{\bf{0}}$,
\end{itemize}
then
\begin{itemize}
\item[{\rm I.}]
$\forall t , t \in \{1,2,\cdots,n_{_{0}}\}  $,
$ \state_{_{t+1}} $ depends on the 
$ \ell + t $ previous cipher symbols.
\item[{\rm II.}]
$\forall t,   t \in \{n_{_{0}}+1,n_{_{0}}+2,\cdots\} $,
$ \state_{_{t+1}} $ depends on the 
$ \ell + n_{_{0}} $ previous cipher symbols.
\item[{\rm III.}]
After $n_{_{0}}$ time step, an unknown input observer  can detect correct plain symbols. 
\end{itemize}
\end{alphthm}
{\bf Sketch of proof.}\\

For (I), at first, by induction  on $ t\geq 1 $ we prove an equivalent explicit form of the internal state $ \state_{_{t+1}} $ as follow,
\begin{equation}\label{explicit}
\state_{_{t+1}}=\sum_{_{j=1}}^{^{t}}\Dmat^{^{j-1}}[(\I-\Emat\Fmat^{^{-1}})\Wmat\memvec_{_{t-j+1}}+\Emat\Fmat^{^{-1}}\ciphervec_{_{t-j+1}}]+\Dmat^{^{t}}\state_{_{1}}.
\end{equation}
Now by definition \ref{mem dfn} of $ \memvec_{_{t}} $ we have,
\begin{center}
$\memvec_{_{t}}= [\mem_{_{t}}^{^{(\len)}},\mem_{_{t}}^{^{(\len -1)}}, \cdots, \mem_{_{t}}^{^{(2)}}, \mem_{_{t}}^{^{(1)}}]^{^{T}} \isdef [\cipher_{_{t-1}}^{^{(1)}},\cipher_{_{t-2}}^{^{(1)}}, \cdots, \cipher_{_{t-\len}}^{^{(1)}}]^{^{T}} $ 
\end{center}
and consequently, for all $  1 \leq j \leq t $, we can write $ \memvec_{_{t-j+1}} $ as follow,
\begin{center}
$\memvec_{_{t-j+1}}= [\cipher_{_{t-j}}^{^{(1)}},\cipher_{_{t-j-1}}^{^{(1)}}, \cdots, \cipher_{_{t-j+1-\len}}^{^{(1)}}]^{^{T}}, $ 
\end{center}
proving part (I).\\

For (II),  suppose that there exist $ n_{_{0}} \in \Integer $ such that $ \Dmat^{^{n_{_{0}}}}=\mathbf{\bf{0}}$. 
Then expand $ \state_{_{t+1}} $ for $t = n_{_{0}}+1 $ as,
\begin{center}
$\begin{array}{lllllll}
\state_{_{t+1}}=&
\Dmat^{^{0}}[(\I-\Emat\Fmat^{^{-1}})\Wmat\memvec_{_{n_{_{0}}+1}}+\Emat\Fmat^{^{-1}}\ciphervec_{_{n_{_{0}}+1}}]+ \\
&\Dmat^{^{1}}[(\I-\Emat\Fmat^{^{-1}})\Wmat\memvec_{_{n_{_{0}}}}+\Emat\Fmat^{^{-1}}\ciphervec_{_{n_{_{0}}}}]+\\
& \vdots \\
&\Dmat^{^{n_{_{0}}-1}}[(\I-\Emat\Fmat^{^{-1}})\Wmat\memvec_{_{2}}+\Emat\Fmat^{^{-1}}\ciphervec_{_{2}}]+ \\
&\Dmat^{^{n_{_{0}}}}[(\I-\Emat\Fmat^{^{-1}})\Wmat\memvec_{_{1}}+\Emat\Fmat^{^{-1}}\ciphervec_{_{1}}]+ \\
&\Dmat^{^{n_{_{0}}+1}}\state_{_{1}},
\end{array}$
\end{center}

and since $ \Dmat^{^{n_{_{0}}}}=\mathbf{\bf{0}}$,  for any $ \nu \geq1 $ and an arbitrary state $ \state_{_{n_{_{0}}+\nu}} $ 
we have,
\begin{center}
\begin{equation}\label{phi-2}
\begin{array}{llllll}
\state_{_{n_{_{0}}+\nu}}&=
\Dmat^{^{0}}[(\I-\Emat\Fmat^{^{-1}})\Wmat\memvec_{_{n_{_{0}}+\nu-1}}+\Emat\Fmat^{^{-1}}\ciphervec_{_{n_{_{0}}+\nu-1}}]+ \\
& \Dmat^{^{1}}[(\I-\Emat\Fmat^{^{-1}})\Wmat\memvec_{_{n_{_{0}}+\nu-2}}+\Emat\Fmat^{^{-1}}\ciphervec_{_{n_{_{0}}+\nu-2}}]+ \\
& \ \ \ \vdots \\
& \Dmat^{^{n_{_{0}}-1}}[(\I-\Emat\Fmat^{^{-1}})\Wmat\memvec_{_{\nu}}+\Emat\Fmat^{^{-1}}\ciphervec_{_{\nu}}]+ \\
&\isdef\phi_{_{2}}(\ciphervec_{_{\nu-\len}}^{^{(1)}},\cdots,\ciphervec_{_{n_{_{0}}+\nu-1}}^{^{(1)}}), \\
\end{array}
\end{equation}
\end{center}

proving (II).

For (III), define $ \errorve_{_{t+1}} \isdef \state_{_{t+1}}-\widehat{\state}_{_{t+1}} $ and note that by \ref{kernel-yek3-1} and \ref{kernel-yek4-1} we have,
\begin{equation}\label{yek-err-1}
\begin{array}{llll}
\errorve_{_{t+1}}&=\state_{_{t+1}}-\widehat{\state}_{_{t+1}}\\
&=\Wmat\memvec_{_{t}}+\Dmat \state_{_{t}}+\Amat\BigPi_{_{\key}}(\state_{_{t}})+\Emat\BigPi_{_{\key}}(\plainvec_{_{t}})-\\
&(\Wmat\memvec_{_{t}}+\Dmat \widehat{\state}_{_{t}}+\Amat\BigPi_{_{\key}}(\widehat{\state}_{_{t}})+\Emat\Fmat^{^{-1}}(\ciphervec_{_{t}}-\widehat{\keystreamvec}_{_{t}})) \\
&=\Dmat \errorve_{_{t}}+\Amat\BigPi_{_{\key}}(\state_{_{t}})+\Emat\BigPi_{_{\key}}(\plainvec_{_{t}})-\\
&\Amat\BigPi_{_{\key}}(\widehat{\state}_{_{t}})-\Emat\Fmat^{^{-1}}(\ciphervec_{_{t}}-\widehat{\keystreamvec}_{_{t}}).\\
\end{array}
\end{equation}

Using  \ref{kernel-yek3-1} and  \ref{kernel-yek4-1} one may conclude that,
\begin{equation}\label{yek-err-3}
\begin{array}{llll}
\errorve_{_{t+1}}&=\Dmat \errorve_{_{t}}+\Amat\BigPi_{_{\key}}(\state_{_{t}})-
\Emat\Fmat^{^{-1}}(\Wmat\memvec_{_{t}}+\Bmat\BigPi_{_{\key}}(\state_{_{t}})-\Wmat\memvec_{_{t}})\\
&=\Dmat \errorve_{_{t}}+\Amat\BigPi_{_{\key}}(\state_{_{t}})-
\Emat\Fmat^{^{-1}}(\Bmat\BigPi_{_{\key}}(\state_{_{t}}),\\
\end{array}
\end{equation}

and consequently, 
\begin{equation}\label{yek-err-4}
\begin{array}{llll}
\errorve_{_{t+1}}&=\Dmat \errorve_{_{t}},\\
\end{array}
\end{equation} 

that proves (III).

\hfill $\blacksquare$

\section{Performance analysis} \label{analysis}

In this section, we concentrate on the performance and statistical evaluation of our proposed scheme 
\cpsc. 

First, let us consider the performance of the scheme in general and most importantly in lightweight setups.
Since, to the best of our knowledge, there is no self-synchronous image encryption scheme similar to our proposed scheme,
we have decided to compare our scheme with Moustique \cite{DK07} which is one of the fastest proposed self-synchronous stream cipher existing so far\footnote{Although there exists severe attacks to Moustique (e.g. see \cite{KRBRRS08}), we have chosen 
this scheme since we are not aware of any better self-synchronized stream cipher similar to what we have proposed.} \cite{GB08}.

In this regard, consider a $4N$-bit input plaintext given to both systems. To generate the ciphertext, the number of field operations for  Moustique is $6000N$. On the other hand, for \cpsc we may consider two sets of parameters which are comparable with 
Moustique, namely ($prec=16$, $\ell=8$, $n=6$ and $\Field_{_{q}}=GF(16)$) with key length $97$ and 
($prec=32$, $\ell=8$, $n=5$ and $\Field_{_{q}}=GF(16)$) with key length $119$. 
In both of these setups the number of field operations to produce the ciphertext given a $4N$-bit plaintext is 
$528N$ which shows that \cpsc is about $11$ times faster than Moustique in bit production. Of course one should also 
note that in our setup we use a bandwidth $8$ times more than Moustique, which give rise to an over-all speed factor of $1.5$ in favor of \cpsc.

\begin{figure}[ht]
\centerline{\includegraphics[width=7cm]{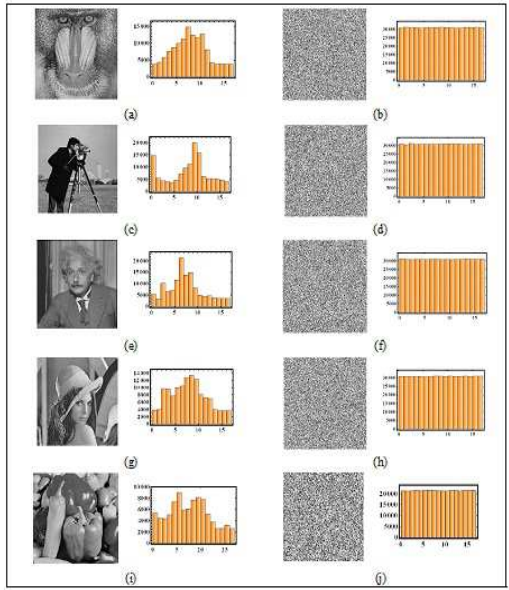}}
\caption{\label{fig:Images and Histogram Table }   
 Images (a),(c),(e),(g),(i) depict histograms of original images  and images (b),(d),(f),(h),(j) depict
  histograms of encrypted images.
}
\end{figure}

\begin{figure}[ht]
\centerline{\includegraphics[width=5cm]{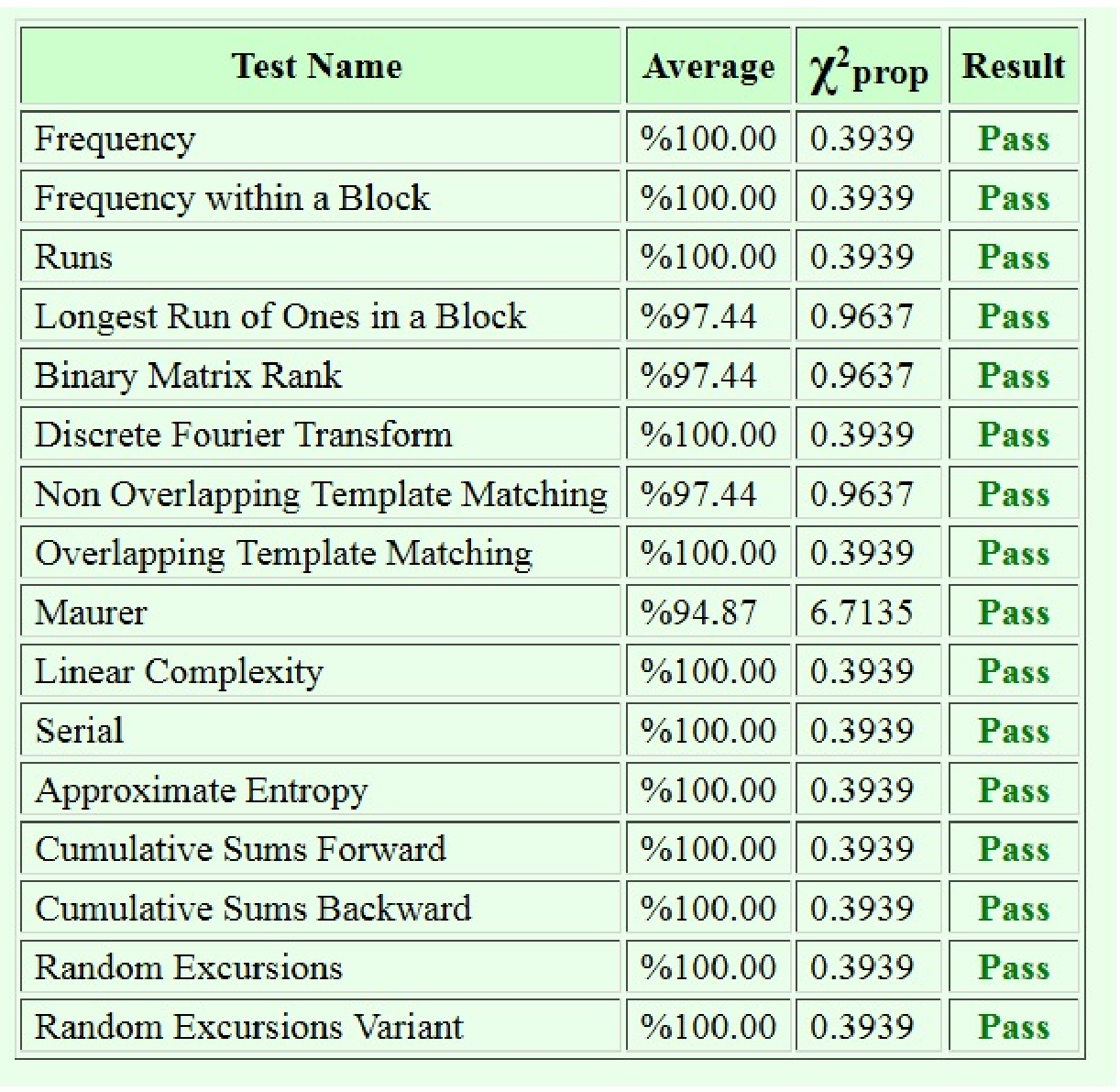}}
\caption{\label{stat Analysis}   \  Randomness of binary sequences for Peper cipher image  }
\end{figure}

To make sure about the uniformity of the output distribution first refer to 
Figure~\ref{fig:Images and Histogram Table } that shows the histograms of some original standard gray images and their corresponding encrypted images, showing an almost uniformly distributed outputs.
In order to be more precise, we have used  NIST Sp-800 Suite \cite{NIST10}  tests for $40$ binary sequences of cipher 
images with $1000,000$ bit length, generated for different secret keys. As it is reported in Figure~\ref{stat Analysis}
for the Peper image below,
\cpsc passes all these tests with an acceptable confidence interval.

On the other hand, in order to test the influence of changing a single symbol in the original image on the encrypted image, the number of symbols' change rate is measured by calculating NPCR (number of symbol change rate) and UACI (unified average changing intensity) as follows (e.g. see \cite{SZJ12} for more on these standard parameters),
\begin{center}\label{NPCR}
$NPCR=\dfrac{100 }{W\times H}\sum_{_{i,j}}D(i,j) \ \ \ ,\ \ \ UACI=\dfrac{100 }{W\times H}\sum_{_{i,j}}\dfrac{\mid C(i,j)-C'(i,j)\mid}{q-1}$
\end{center}
in which $W$ and $H$ are the width and the height of encrypted images. Note that NPCR measures the percentage of different symbols  between the two cipher images and UACI measures the average intensity of differences between the two cipher images. Two encrypted images $C$ and $C'$, whose corresponding original images $P$ and $P'$  have only one-symbol difference, are considered. A two-dimensional array $D$ with the same size of $C$ and $C'$  is defined for which $D(i,j) = 1$  if $C(i,j) \not = C' (i,j)$, and $D(i,j) = 0$ otherwise. As Table~\ref{tab: sensitivity analysis} reflects our experimental results, \cpsc has good cipher image sensibility to little purtubation in plain images.
\begin{table}[!b]
\renewcommand{\arraystretch}{1.0}
\caption{Plain image sensitivity analysis}
\centering
\label{tab: sensitivity analysis}
\begin{tabular}{|c|c|c|c|c|c|}
\hline$   $ & $ NPCR $& $ UACI$ \\
\hline 
\hline Baboon & 99.552  & 33.171   \\
\hline Camera Man & 99.572   & 33.239  \\
\hline Einstein &99.543 & 33.229   \\
\hline Lena & 99.540   & 33.239  \\
\hline Peper &99.597&33.250\\
\hline
\end{tabular}
\end{table}

Also, the cipher image dependency on a small perturbation in secret key bits is analyzed. Table~\ref{tab: key sensitivity analysis} shows the detailed results for the encryption of the Peper image with two secret keys, which have just a one bit difference. 

\begin{table}[!t]
\renewcommand{\arraystretch}{1.0}
\caption{Key sensitivity analysis}
\centering
\label{tab: key sensitivity analysis}
\begin{tabular}{|c|c|c|c|c|c|}
\hline$   $ & $ NPCR $& $ UACI $  \\
\hline 
\hline Peper & 99.549&33.219 \\
\hline
\end{tabular}
\end{table}
\subsection{Cipher image entropy and correlation analysis} \label{entro anal}
Information entropy is one of the most significant features of randomness. Information entropy $h(m)$ of a message $m$ can be measured by the following formula,

\begin{equation}
h(m)=-\sum_{_{i=0}}^{^{n-1}}p(m_{_{i}})log_{_{2}}(p(m_{_{i}})),
\end{equation}
where $  n$ is the total number of symbols in the message $m$ and  $p(m_{_{i}}) $ represents the probability of the occurrence of symbol $ m_{_{i}} $.
Theoretically, for a random code source with an alphabet of size $n$, the ideal information entropy must be  $ h(m)=\log_{_{2}}(n)$. 
Table \ref{tab:Entropy analysis} shows that in \cpsc the entropy of encrypted images for four standard images are close to  ideal values, which shows robustness against entropy attacks.

\begin{table}[!b]
\renewcommand{\arraystretch}{0.80}
\caption{Byte, symbol and bit entropy analysis of plain and cipher images}
\centering
\label{tab:Entropy analysis}
\begin{tabular}{|c|c|c|c|c|c|c|c|c|}
\hline$   $ & $ Bit \ entropy $ &  & $ Symbol \ entropy $&   & $ Byte \ entropy $&\\
\hline$   $ & $ Plain $& $ Cipher $ & $ Plain $& $ Cipher $& $ Plain $& $ Cipher $\\
\hline 
\hline Baboon &  0.84 & 0.97 & 3.92 &  4.0 & 7.33 & 7.69 \\
\hline Camera Man & 0.77  &0.97  & 3.85 & 4.0  & 7.01 & 7.69\\
\hline Einstein& 0.78  & 0.97 & 3.83 & 4.0  &6.88  & 7.68\\
\hline Lena & 0.79  & 0.97 & 3.94 & 4.0  & 7.44 &7.68 \\
\hline Peper & 0.77  & 0.97 & 3.98 & 4.0  & 7.59 &7.69 \\
\hline
\end{tabular}
\end{table}

Also, there is usually strong correlations between adjacent symbols in the input image. A secure image encryption scheme should remove this correlation to make statistical attacks infeasible. In order to test the correlation between adjacent symbols, $2500$ random pairs of adjacent symbols (in horizontal, vertical, and diagonal directions) are selected and the correlation coefficient of each pair is computed before and after encryption using the following equations,

\begin{figure}[h]
\centerline{\includegraphics[width=5cm]{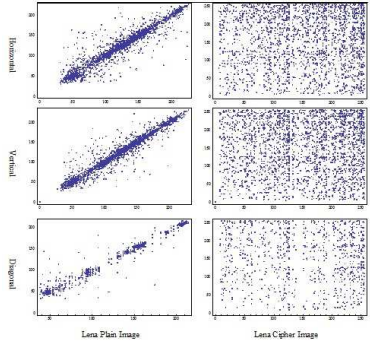}}
\caption{\label{fig:Images Correlation Table}   
 \ Correlation plot of  adjacent symbols in the Lena plain and cipher images}
\end{figure}

\begin{equation}\label{corr}
Cor=\dfrac{\sum_{_{i=1}}^{^{N}}(x_{_{i}}-\bar{x})(y_{_{i}}-\bar{y})}{\sqrt{\left( \sum_{_{i=1}}^{^{N}}(x_{_{i}}-\bar{x})^{^{2}})\right) \left( \sum_{_{i=1}}^{^{N}}(y_{_{i}}-\bar{y})^{^{2}}\right) }}
\end{equation}
where   $ \bar{x} $  and $ \bar{y} $ are average values.
The correlation plot of the plain image and the cipher image of Lena is illustrated in Figure~\ref{fig:Images Correlation Table}. Also,  Table \ref{tab:Correlation analysis} summarizes the results corresponding to $4$ other images.
\begin{table}[!b]
\renewcommand{\arraystretch}{0.80}
\caption{Correlation analysis of plain and cipher images}
\centering
\label{tab:Correlation analysis}
\begin{tabular}{|c|c|c|c|c|c|c|c|}
\hline$   $ && $ Plain  $ &&&   $Cipher $ &\\
\hline$   $ & $ Hori. $& $ Vert. $ & $ Diag.$& $ Hori. $& $ Vert. $& $ Diag. $\\
\hline 
\hline Baboon & 0.669 & 0.723 & 0.643 &  0.019 & 0.034 & 0.005 \\
\hline Camera Man & 0.935  &0.976  & 0.913 & 0.006  & 0.011 &0.019\\
\hline Einstein& 0.892  & 0.723 &0.912 & 0.027  &0.034  &0.006\\
\hline Lena & 0.910  &0.961 & 0.913 & 0.035  & 0.019 &0.001 \\
\hline
\end{tabular}
\end{table}

\subsection{Self-synchronizing property analysis}
\begin{figure}[h]
\centerline{\includegraphics[width=5cm]{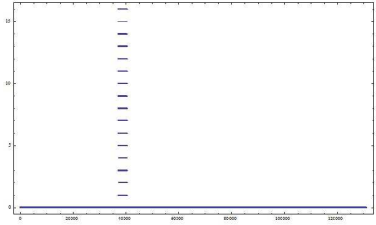}}
\caption{\label{fig:diff ss property}   
 \ Numerical difference of sent cipher image data  and received the one. }
\end{figure}

\begin{figure}[h]
\centerline{\includegraphics[width=5cm]{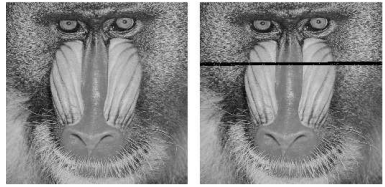}}
\caption{\label{fig:ss property}   
 Error recovery in the self-synchronous mode for Baboon.}
\end{figure}

Another feature of \cpsc is the fact that the scheme can be used in a self-synchronous mode maintaining error correction 
 using an unknown input observer as a receiver (e.g. see \cite{Mil08} for more on this method). To show this property, 
Figures~\ref{fig:diff ss property} and \ref{fig:ss property} demonstrate the result of and error recovery in the Baboon image where the experiment are depicted numerically and graphically in these figures.

\end{document}